\documentstyle[11pt,newpasp,twoside]{article}
\def\edcomment#1{\iffalse\marginpar{\raggedright\sl#1\/}\else\relax\fi}
\reversemarginpar

\begin{document}
\title{Chandra Studies of Star Forming Regions}
\author{Eric D. Feigelson}
\affil{Dept.\ of Astronomy \& Astrophysics, Pennsylvania State
University, University Park PA 16802 USA}

\begin{abstract}
When observed with sufficiently high spatial resolution and
sensitivity, star formation regions are unusually complex X-ray
sources.  Low-mass protostars and T Tauri stars, Herbig Ae/Be stars, OB
and Wolf-Rayet stars are seen at levels $28 \leq \log L_x \leq 34$
erg/s.  High-amplitude variability from magnetic reconnection flares
are often present.  From past star formation episodes, supernova
remnants and X-ray binary systems may dominate the emission on large
scales.  Astrophysically, ionization of molecular material from yount
stellar X-rays which may have important consequences for circumstellar
disk evolution, bipolar flow ejection and star formation.

We report here early results from Chandra X-ray Observatory studies of
a wide range of star forming regions using the ACIS camera.  They
include portions of the nearby Ophiuchus and Perseus clouds, the Orion
Nebula Cluster (ONC) and elsewhere in the Orion giant molecular
clouds, molecular clouds and star clusters around the Galactic
Center, the 30 Doradus region of the LMC, and the prototype
starburst galaxy M 82.  

\end{abstract}

\section{Introduction}

The astrophysics of star formation and the early phases of stellar
evolution have proved to be very complex phenomena that are currently
under intense study.  The traditional  model of  collapse of a uniform
molecular cloud core when the gravitational force exceeds the thermal
gas pressure (the Jeans criterion) is far too simple.  Turbulent and
magnetic pressures are comparable to thermal pressures in cloud cores,
and the cores themselves race around supersonically within the giant
molecular clouds (GMCs).

Even after loss of angular momentum via fragmentation, a significant
fraction of the collapsing material enters a circumstellar disk rather
than the protostar.  The star-disk system drives powerful outflows,
both highly collimated high-velocity Herbig-Haro objects and broader
slower molecular bipolar flows.  At the same time, material accretes
onto the star, most likely funneled along magnetic field lines and the
star contracts along the pre-main sequence convective Hayashi tracks
towards the Zero-Age Main Sequence.  The formation of high mass OB
stars is enigmatic, as the Eddington limit should inhibit their
formation from direct gravitational collapse.  The recent reviews in
{\it Protostars and Planets IV} (Mannings et al.\ 2000) and the
monograph by Hartmann (1998) describe our current understanding of
these issues.

Initially, there seems to be little role in the star formation story
for high energy processes:  the molecular clouds have temperatures $ T
\sim 10$ K, the young stellar surfaces have bolometric  $T \sim
2000-3000$ K with disk temperatures roughly in the range $T \sim 100-
1000$ K.  Galactic cosmic rays are thought to provide a low uniform
level of ionization through molecular clouds, but it has not been
clearly established that most can penetrate the magnetic fields of the
cloud and outflows.

It was thus a surprise in the early 1980s when the $Einstein$ imaging
X-ray telescope first pointed towards nearby star forming regions and
found many hard, variable X-ray sources associated with young stars.
Studies with $ROSAT$ and $ASCA$ in the 1990s found hundreds of young
stars within star forming regions, and thousands of somewhat older
pre-main sequence stars across the sky.  It was established that T
Tauri stars (ages $t \simeq 1-20$ Myr) all emitted X- rays at levels
$10^1-10^4$ times those seen in normal old main sequence stars like the
Sun, and that at least a few protostars ($t \simeq 0.1-1$ Myr) also
emit at very high levels.  X-ray flares, typically lasting hours,
similar but more powerful than those seen in the Sun and dMe flare
stars, are often present.  These X-ray findings and related
astrophysical issues are reviewed by Feigelson \& Montmerle (1999,
henceforth FM99).

Most of the past X-ray studies (and closely related multiwavelength
observations of magnetic activity and flares in PMS stars) have been
made in the nearest, rather small star forming molecular clouds at
distances $d \simeq 150-450$ pc.  Only small glimpses of star formation
regions were possible in the truly giant cloud complexes at distances
$d \simeq 2$ kpc,  across the Galaxy and in nearby galaxies like the
LMC.  Demographics of star forming regions indicates that most stars
form in such rich stellar clusters from GMCs.

The {\it Chandra X-ray Observatory's} high spatial resolution and
sensitivity to hard photons, which penetrate both the local molecular
cloud and intervening spiral arms, provide an enormous increase in
capability for detailed study of star forming regions, particularly
those beyond the solar neighborhood.

\section{Expected X-ray emission from star forming regions}

Figure 1 is a cartoon of a typical GMC from the X-ray point of view.
Three typical situations are shown: an optically visible OB association
at the edge of the GMC producing a blister H {\sc II} region;  an
embedded OB association seen in K-band or mid-infrared observations
producing (ultra-)compact H {\sc II} regions; and distributed star
formation.  The diagram is not to scale: the GMC is an inhomogeneous
and probably turbulent structure  $\simeq 100$ pc in size, while the young
star clusters are typically $1-2$ pc in size.  The X-ray components of a 
star forming region can be categorized as follows:

\begin{figure}[!ht]
\plotfiddle{"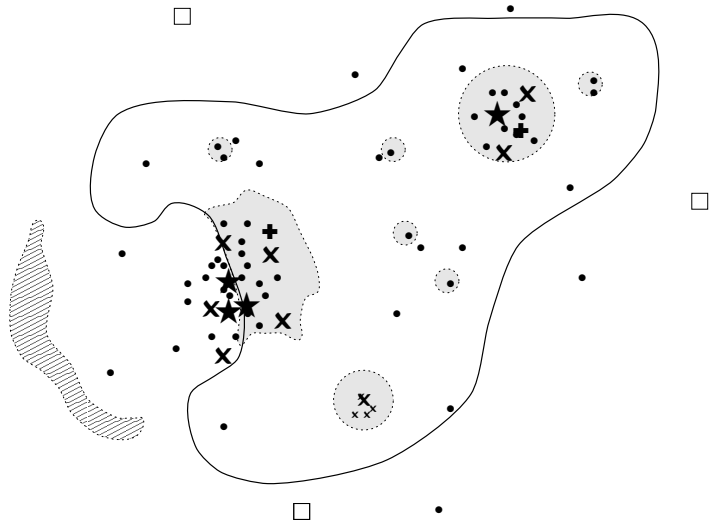"}{3.2in}{0.0}{100.}{100.}{-150.0}{0.0}
\caption{Diagram of the expected X-ray components from a giant
molecular cloud with a blister H{\sc II} region, embedded young star
cluster and distributed star formation.  Symbols: {\Large\bf $\star$} = 
OB
stars; {\bf $\times$} = Herbig Ae/Be stars; $\bullet$ = T Tauri stars;
{\bf +} = protostars; $\Box$ = X-ray binary system.  Hatched
region outside of the cloud represents a supernova remnant, and shaded
regions within the cloud represent partially ionized X-ray dissociation
regions.}
\end{figure}

{\bf T Tauri stars} ~~ The most numerous X-ray source type found in
star formation regions are low-mass PMS stars ($M \leq 2$ M$_\odot$)
with a broad X-ray luminosity function ranging from $< 10^{28}$ erg/s
for the least massive brown dwarfs to $10^{31}$ erg/s for the most
luminous massive members of the class.  The X-ray are produced both
from classical T Tauri stars with star-disk accretion and ejection and
from weak-lined T Tauri stars where the disk is largely dissipated.
Flares with timescales of hours are often seen in these X-ray sources.
The astrophysical origin is readily attributable to solar-type magnetic
activity: plasma heated to $10^7$ K temperatures in multipolar magnetic
field loops rooted onto the stellar surface (FM99).

{\bf Protostars} ~~ Prior to Chandra, only a small fraction of
protostars were detected in X-rays, but these often exhibited unusual
properties:  X-ray temperatures approaching $10^8$ K, some very
powerful flares, and in one case, an iron emission line possibly
indicating fluorescence off of a cold disk.  Here it is possible that
the magnetic reconnection involves long field lines connecting the
protostars to the disk (FM99).

{\bf Herbig Ae/Be stars} ~~  These  intermediate mass pre-main sequence
stars with star-disk interactions show X-ray emission ranging in the
range $10^{29}-10^{31}$ erg/s.  Recent $ASCA$ observations have
detected flares supporting a magnetic reconnection origin as in lower
mass stars (Hamaguchi et al.\ 1999).

{\bf OB and Wolf-Rayet stars} ~~ These massive stars typically emit
X-rays in the $10^{32}-10^{34}$ erg/s range, and the emission is
attributable to shocks in the line-driven wind, as an approximate
relationship $L_x \propto L_{bol} \propto L_{wind}$.  Extremely massive
stars ($M>50$ M$_\odot$) evolve so quickly that post-main sequence
supergiants,including Wolf-Rayet stars, can coexist with pre-main
sequence stars in a single rich young stellar cluster.  X-ray spectra
are generally soft, except in close binaries of very massive stars
where the colliding winds shocks attain $10^8$ K.

{\bf Supernova remnants} ~~  Massive supergiants quickly undergo
supernova explosions and the resulting ejected remnants which are
strong X-ray emitters for $\sim 10^5$ yr.  This coexistence of stellar
birth and death generally occurs only in the most massive star forming
regions; for example, in the discussion below it is seen in 30 Doradus
but not in Orion.  Supernova remnant X-ray luminosities usually lie in
the range $L_x \simeq 10^{35}-10^{37}$ erg/s with spatial scales
around $10^{-1}-10^{1}$ pc.

{\bf X-ray binary systems} ~~ X-ray binaries are close binary systems
where one member is a compact stellar remnant accreting from its
companion via Roche lobe overflow or a stellar wind. While the spatial
association between star forming regions and X-ray binaries may be
tenuous due to the delay in forming accreting systems, X-ray binaries
can outshine all other X-ray components with $L_x \simeq
10^{36}-10^{40}$ erg/s.  Helfand and Moran (2001) have 
estimated that $2-5$\% of OB stars eventually produce a high-mass X-ray
binary, $0.1-0.2$\% are currently in that phase, and the X-ray binary
luminosity is $2-20 \times 10^{34}$ erg/s per O star.

\section{Early Chandra observations}

I outline briefly some of the studies underway during the first
year of Chandra operations, starting with nearby molecular clouds
producing small clusters of lower mass stars and ending with galactic
scale starbursts dominated by massive star formation.  As this review
is being written in November 2000,  many results are still in a
preliminary stage of analysis and interpretation.  

\subsection{$\rho$ Ophichus cloud}

The $\rho$ Ophiuchi cloud  is one of several molecular cloud complexes
lying $\simeq 150$ pc from the Sun on the periphery of the Local Hot
Bubble.  It consists of several dense cloud cores about 0.2 pc in size
now actively forming stars lying at the head of a cometary cloud
complex about 10 pc in extent.  The star formation may
have been triggered by the O-star winds of the nearby Sco-Cen OB
association.

In the first year, Chandra is obtaining two deep $\sim 100$ ks ACIS
exposures of the Core A and Cores E/F regions, and a mosaic of seven
shallow $\sim 5$ ks exposures of outer cloud.  In the mosaic,
about 100 sources, mainly from classical and weak-lines T
Tauri stars (Grosso et al., in preparation).

In the deep image of Cores E/F, 87 sources are detected; for
comparison, the ROSAT HRI detected only 8 sources in the same region
(Iminishi et al.\ 2001).  In addition to T Tauri stars, a remarkably
high fraction of Class I protostars are detected: 15 out of 21 in the
field.  Thus in a single image, Chandra provides more X-ray detections
than obtained from the entire ROSAT and ASCA missions combined!  This
also demonstrates that elevated X-ray emission is a generic property of
low mass stars in their Class I phase, and is not restricted to a few
unusual systems.  Ten Class I sources exhibited X- ray flares
during the observation, clearly demonstrating that violent magnetic
reconnection events produce protostellar X-ray emission.
Finally, the X-ray spectrum of Class I protostar YLW 16A shows a strong
fluorescent Fe 6.4 keV line, directly demonstrating that X-rays will
shine upon and ionize the circumstellar disk (see \S 5).

\subsection{Perseus molecular cloud}

Chandra has scheduled  observations of the two rich star clusters in
the Perseus molecular cloud ($d \simeq 350$ pc): IC 348 with stars
several Myr old, and NGC 1333 with stars $\leq 1$ Myr old.  The
protostars of NGC 1333 produce over a dozen powerful bipolar flows
which are violently stirring up the cloud.  Preliminary examination of
the 50 ks ACIS exposure of NGC 1333 shows over 80 X-ray sources, most
associated with embedded members of the star cluster known from K-band
studies (Feigelson et al., in preparation).  Spectral types of X-ray
detected soruces range from mid-M to mid- O.  The pattern of X-ray
emission is not always straightforward:   for example, the M4 star ASR
7 with $A_V \simeq 20$ shows more X-ray counts than the nearby O6 star
ASF 8 with $A_V \simeq 7$.   In contrast with the $\rho$ Ophiuchi
cloud, most of the Classs 0 and I protostars in the NGC 1333 field are
not detected,

\begin{figure}
\plotfiddle{"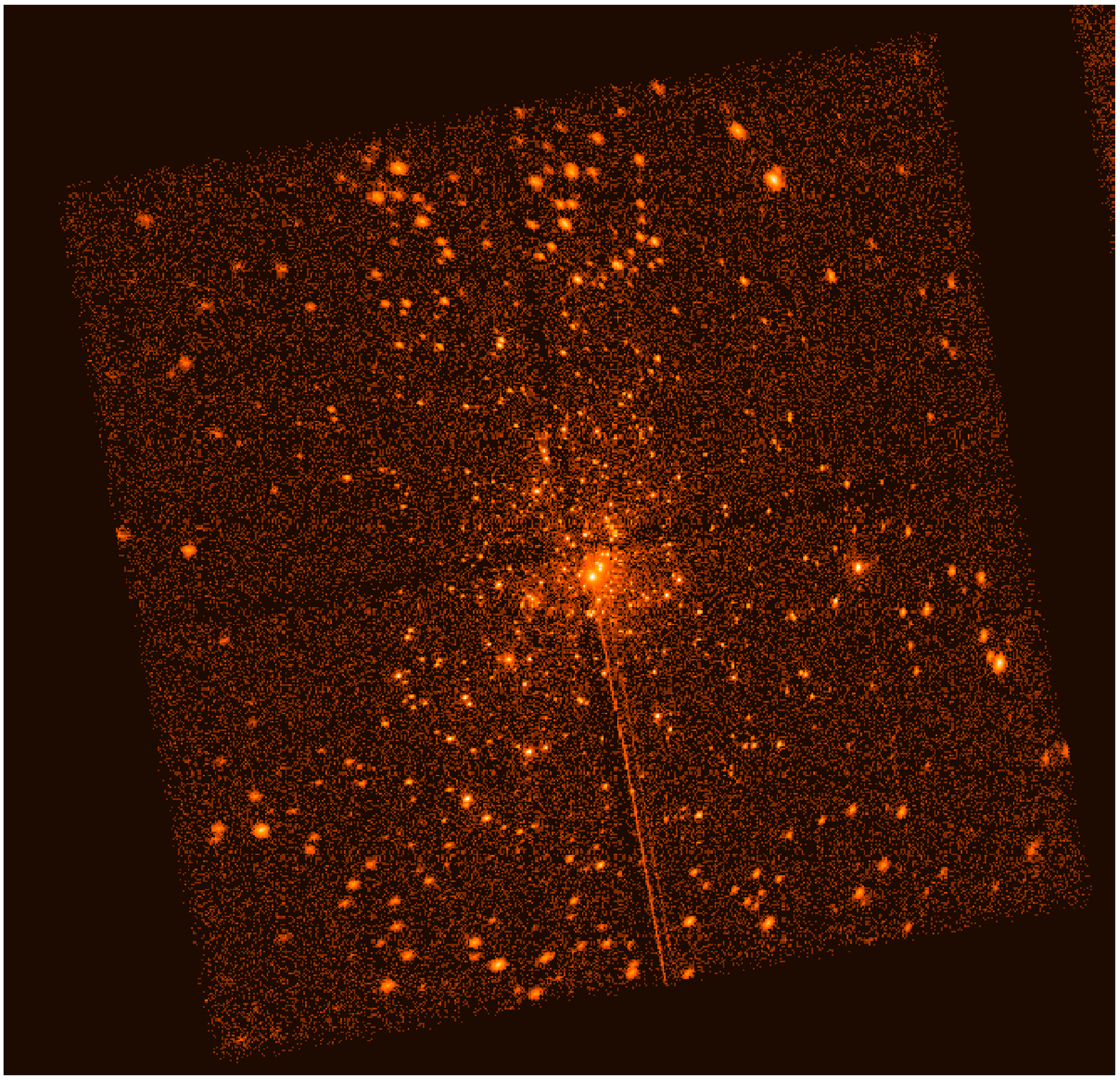"}{4.0in}{0.0}{40.}{40.}{-250.}{50.}
\plotfiddle{"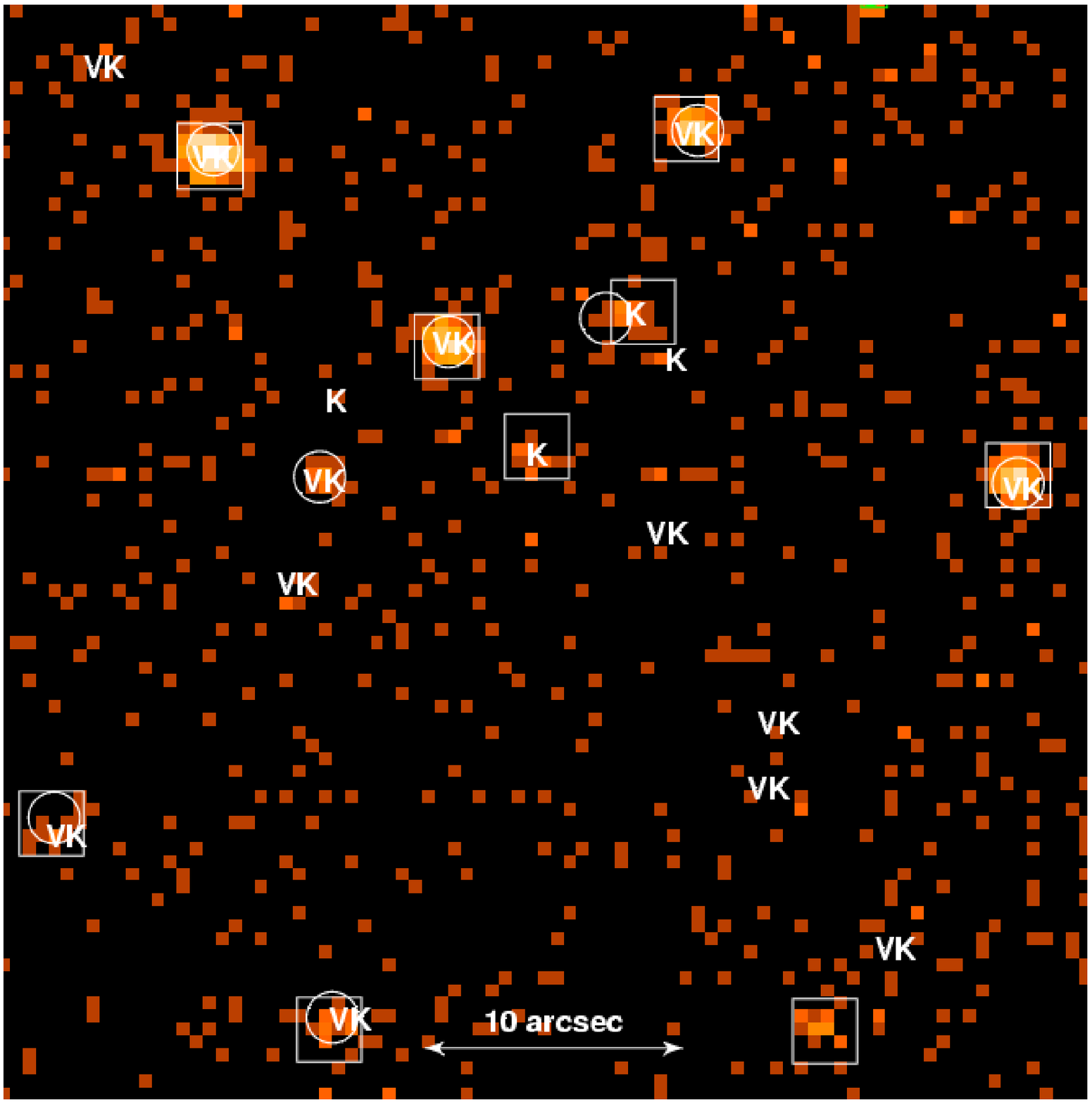"}{3.9in}{0.0}{39.}{39.}{0.}{350.}
\vspace*{-4.7in}
\caption{ACIS image of the Orion Nebula region.  {\it Left:} The entire
field ($18^\prime \times 18^\prime$) with $>$1000 sources clustered
around the Trapezium OB stars.  {\it Right:} A small region of this
field showing the low background and high resolution of the
instrument.  Symbols: $\bigcirc$ = soft X-ray source (usually
unobscured T Tauri stars in the H{\sc II} region); $\Box$ = hard X-ray
source (embedded stars); V = optical brighter than V=20; and K =
infrared brighter than K=18.  Only one source in this region was
previously detected with ROSAT. From Garmire et al.\ (2000)}
\end{figure}

\subsection{Orion molecular cloud complex}

The early  Chandra program includes  intense study of the nearest giant
molecular clouds at $d \simeq 450-500$ pc, including several deep
observations of the Orion Nebula region and several exposures elsewhere
in the Orion A and B clouds.

The Chandra images of the ONC are spectacular with over a thousand
sources detected in a single field --  a record for X-ray astronomy.
The technical achievements of thes observations are truly impressive:
sources are detected as faint as $\simeq 6$ photons (limiting $L_x
\simeq 1 \times 10^{28}$ erg/s, better than the most sensitive ROSAT
study of even the nearest star forming regions), stars detected
embedded up to $A_V \simeq 60$ magnitudes ($N_H \simeq 1 \times
10^{23}$ cm$^{-2}$), binaries resolved on-axis as close as
$2^{\prime\prime}$ (1000 AU), and absolute astrometry achieved to $\pm
0.1^{\prime\prime}$ on the Hipparcos reference frame (Garmire et
al.\ 2000).  About half of the ONC cluster members are detected with a
strong mass dependence:  virtually all of members above 1 M$_\odot$,
stars between $0.2-1$ M$_\odot$ if the absorption is not too high, and
very few of the very low mass stars and brown dwarfs.

A few of many specific results expected from these observations have
been described (Garmire et al.\ 2000, Schulz et al.\ 2001a).   A wide
range of variability behavior is seen.  A few sources are invisible
except during flares lasting several hours, and there may be a trend
towards high variability for stars with higher absorption.   X-rays are
seen from young stars both with and without circumstellar disks,
including those embedded in ionized globules seen in HST images.  A
considerable fraction of the known radio-emitting young stars in the
ONC are detected; in most cases, the radio continuum is non-thermal
gyrosynchrotron emission associated with powerful magnetic flares.
Examination of the evolution of X-ray luminosity among a subsample of
solar-mass stars suggests a previously unreported effect:  emission
appears uniformly high ($L_x \simeq 2 \times 120^{30}$ erg/s) during
the first $1-2$ Myr, but diverges between high and low emitters at ages
around $2-10$ Myr.  This may be explained by theories of the
regulation of stellar angular momentum by magnetic coupling with
circumstellar disks.

Observations of the ONC Trapezium stars provide a new detailed view of
OB X-ray emission (Schulz et al.\ 2001a and 2001b).  The most massive
members, $\theta^1$ Ori A, C and E with spectral types O7$-$B0.5 have
$L_x  \simeq 2-20 \times 10^{31}$ erg/s, while the less massive
$\theta^1$ Ori B and D with spectral type B1 emit only $L_x \simeq 5-10
\times 10^{29}$ erg/s, comparable to typical lower-mass T Taur stars.
Chandra grating observations of the most luminous $\theta^1$ Ori C
component reveals 30 lines from six elements (O, Ne, Mg, Si, S and Fe)
with line ratios indicating a wide range of temperatures ($0.5-60$ MK)
and densities ($0.3-9 \times 10^{13}$ cm$^{-3}$) Many lines are resolved
with Doppler widths ranging from $\sim 400$ to 2000 km/s with a trend
of higher velocities for the higher energy lines.  The temperatures are
considerably higher than predicted from the long-standing model of
instabilities in a line-driven wind -- emission from thin dense shells
far out into the wind is a more promising approach.

In addition to the astrophysical insights likely to emerge from the new
spectroscopy of young OB stars, the astronomical implications are
clear: if OB stars typically have significant fractions of their hot
plasma at energies in excess of $2-3$ keV, then these stars will be
detectable even through the densest molecular cloud interiors and
intervening Galactic obscuration.  The $2-6$ keV bandpass is comparable
to the mid- to far- infrared bands in its ability to penetrate high
column densities.  Chandra and XMM can thus study massive star
formation throughout much of the Galactic disk.  Early indications of
this can be seen in the Orion Nebula field: mid-infrared Source n, one
of the BN/KL cluster members, is faintly detected in the ACIS image,
and a deeply embedded source $1.1^{\prime\prime}$ from the BN object
itself are weakly detected in the $2-8$ keV band.  It is not clear
whether the latter is produced somehow by the massive BN object (shocks
in its outflow?) or is due to an unusual previously unknown lower mass
star.  The hard X-ray emission of OB stars may also have important
implications for the ionization of molecular cloud cores (\S 5 below).

Several Chandra observations are being made in the Orion giant
molecular clouds away from the Orion Nebula itself.  A short 2 ks
observation of a field 1$^\circ$ to the south in the Orion A cloud
reveals 18 T Tauri stars with spectral types from A0 to M5 (Pravdo et
al.\ 2001).  Comparison with Einstein and ROSAT observations of the
same field gives a 20-year baseline for variability studies.  This and
similar studies may also help define the ratio of stars formed inside
and outside of rich clusters: X-ray surveys exclusively isolate
pre-main sequence stars, while  K-band observations of unconcentrated
regions of star formation are frequently overwhelmed by background
sources.

A remarkable field has been studied 0.2$^\circ$ north of the Orion
Nebula, pointed at the filament-shaped OMC 2/3 dense molecular cloud
cores.  Millimeter studies have shown this is a site of intense current
star formation with up to 30 likely protostars of Class 0 and I.  In a
deep 100ks ACIS exposure, several deeply embedded X-ray sources are
found around OMC 3, two of which coincide with millimeter Class 0
protostars (Tsuboi et al.\  2001).  This finding may have profound
importance for the astrophysics of protostars and possibly the star
formation process itself.   It strengthens the concept that X-ray
ionization is involved in launching bipolar flows and in promoting
accretion from the circumstellar disk, both of which are strongest in
the Class 0 phase (\S 4).

\subsection{The Galactic Center region}

The inner $\simeq 100$ pc of the Milky Way Galaxy has a profusion of
unusually massive giant molecular clouds with many young OB
associations.  A $2.6 \times 10^6$ M$_\odot$ black hole resides at the
dynamical center, Sgr A*.   If viewed from afar, our Galaxy would
probably be classified as a mild nuclear starburst galaxy with an
extremely weak active galactic nucleus.  The configuration of
molecular, atomic and ionized gas is extremely confused in the Galactic
Center region due both to the increased concentration of matter and the
rapidly changing rotation curve.

The Chandra ACIS image of the innermost 20 pc shows a profusion of 
X-ray binary systems, highly structured dense hot plasma, a bright 
supernova remnant (Sgr A-East) which may be interacting with Sgr A*, 
individual young massive stars, and a very weak source with $L_x \simeq 
2 \times 10^{33}$ erg/s associated with Sgr A* itself (Baganoff et al.\ 
2001).  Several of the resolved sources within 1 pc of the nucleus are 
known Wolf-Rayet stars with estimated masses $>10$ M$_\odot$, ages 
up to 5 Myr, and X-ray luminosities of order $10^{33}$ erg/s.  But the 
stellar X-ray emission is dwarfed by Sgr A-East and the diffuse X-ray 
emission, which may arise from sheared earlier supernova remnants
(Maeda et al.\ 2001). 

About 100 pc from the center lies the Sgr B2 giant molecular cloud.
Chandra has confirmed the remarkable ASCA finding that the cloud is
glowing in fluorescent Fe K$\alpha$ 6.4 keV line (Murakami et al., in
preparation).  This is the first case of an X-ray reflection nebula,
and another case is emerging from the Sgr C molecular cloud (Murakami
et al.\ 2001).  These molecular clouds glowing in X-rays require that
an X-ray source towards the Galactic Center (possibly Sgr A* itself)
was bright $\sim 10^2$ yrs ago but has now dimmed.  The ACIS image also
shows a heavily absorbed faint X-ray source at the center of the cloud
and associated with an IRAS source.  With $N_H \simeq 9 \times 10^{23}$
cm$^{-2}$ ($A_V \simeq 500$), $kT \simeq 5$ keV and $L_x \simeq 2
\times 10^{33}$ erg/s, it is probably a young stellar cluster somewhat
more massive than the Orion Nebula cluster lying on the far side of the
cloud.  This detection demonstrates Chandra's capability of finding
star forming regions in even the most obscured regions of the Galactic
disk.

\begin{figure}[!ht]
\plotfiddle{"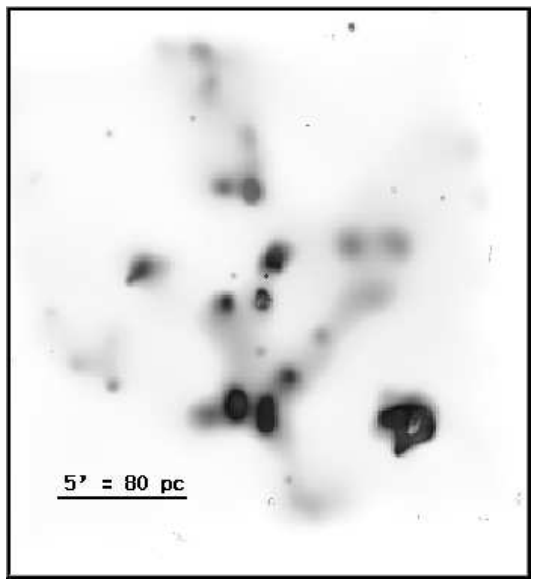"}{2.9in}{0.0}{70.}{70.}{-100.}{0.}
\caption{ACIS-I image of the 30 Doradus region after adaptive
smoothing.  Features include the bright composite supernova remnant
N157B to the southwest, many X-ray binaries, and highly structured
large superbubbles surrounding the R136a young stellar cluster.  R136a
lies within the small central spot, where the emission is dominated by
Wolf-Rayet binary systems. Supernova remnants N157C and 1987a lie to
the southwest on ACIS-S chips not shown here.  From Townsley et al., in
preparation.}

\end{figure}

\subsection{The 30 Doradus region}

30 Doradus in the Large Magellanic Cloud at $d \simeq 50$ kpc, known 
optically as the Tarantula Nebula, is one of the few young `super star 
clusters' in the Local Group of Galaxies with sufficient mass and 
concentration to evolve into a globular cluster.  The cluster has 
$>300$ OB stars, many with masses $>100$ M$_\odot$, both on the main 
sequence and in the Wolf-Rayet phase.  These massive stars have 
produced powerful winds and supernova remnants that have ionized a 
$\simeq 200$ pc region and largely destroyed the parental molecular 
cloud.  

The 25ks ACIS image of 30 Dor is extremely complex (Figure 3, Townsley
et al.\ in preparation; see also Dennerl et al.\ 2001 for an XMM
observation of the region).  The dense young stellar cluster R136a at
the field center appears as a faint $1^{\prime\prime}-2^{\prime\prime}$
emission region with $L_x \simeq 10^{33}$ erg/s.  These many ordinary
OB stars are outshined by a dozen individually resolved O3 and WN stars
which are known or likely close binaries with masses $80-140$ M$_\odot$
lying within 10$^{\prime\prime}$ of the central cluster.  These sources
with individual $L_x \sim 3-30 \times 10^{33}$ erg/s are likely due to
colliding stellar winds.  Scattered on scales of tens of parsecs are
dozens of X-ray binaries each with $L_x \simeq 10^{33}-10^{35}$ erg/s.
Several arcmins from R136a are three remarkable supernova remnants:
N157B, which the Chandra data clearly shows is a composite
shell-plus-plerion remnant with a central bright X-ray pulsar; the
large shell-like N157C remnant; and SN 1987a.

\subsection{Messier 82}

The largest scale of star formation known is the starburst galaxy of 
which M 82 is the closest example at $d \simeq 3.6$ Mpc.  The inner 
kiloparsec of its disk has many super star clusters, each comparable to 
30 Doradus, with a net supernova rate around 0.1/yr.  Most of these 
giant OB associations are deeply embedded in dusty clouds so that the 
reprocessed infrared luminosity exceeds its visual luminosity by a 
factor of $10^2$.  M 82 is a sufficiently bright X-ray source to appear 
in the UHURU source catalogs 30 years ago.  For many years, the nature 
of its X-ray emission has been debated: the balance between emission 
from individual supernova remnants, a global hot interstellar medium 
(known to extend several kpc outward into the halo), OB associations, 
X-ray binaries, and a putative active galactic nucleus was uncertain.  
Chandra is resolving these issues for M 82 and other starburst galaxies 
(Griffiths et al.\ 2000; Kaaret et al.\ 2001; Matsumoto et al.\
2001; Matsushita et al.\ 2001; Ward, this conference).

A 35 ks ACIS image shows that $60-75$\% of the $2-10$ keV emission is 
resolved into compact sources and the remainder arises from diffuse 
emission or unresolved sources (Griffiths et al.\ 2000).  Several of 
the $\sim 20$ unresolved sources have $L_x > 3 \times 10^{38}$ erg/s,  
too luminous for neutron star accretion, are likely black holes 
accreting from the winds of close binary O and Wolf-Rayet companions. 
While few individual radio supernova remnants are seen in the X-ray 
image, the global interstellar medium is extremely dense, hot ($T 
\simeq 40$ MK) and highly structured.  Remarkably, its pressure is 
$10^5$ times that of the hot interstellar medium around the Sun.  This 
interstellar medium probably arises from the early merging of thousand 
of remnants produced by the starburst, and clearly drives the galactic 
wind into the halo and intergalactic space.  The brightest unobscured 
super star cluster, M82-A, with optical $L_V \simeq 10^{42}$ erg/s is 
seen with $L_x \simeq 8 \times 10^{38}$ erg/s.  This emission is 
probably dominated by unresolved X-ray binaries, but the telescope 
resolution is insufficient to establish the contribution of normal 
stars or supernova remnants.

\subsection{Future prospects}

Roughly 1 Ms annually or $\simeq 5$\% of Chandra time is devoted to
studies of star formation and young stars.  In addition to those
outlined above, the program of the first two years includes the
following observations:

{\bf Nearby low-mass star forming regions} ~~ R Corona Australis cloud;
Lynds 1551 cloud in Taurus; Core A of the Ophiuchus cloud; northern
portion of the Chamaeleon I cloud; high-latitude MBM 12 cloud; IC 348
in the Perseus cloud; Herbig-Haro 1-2 region in Orion; TW Hya stars;
isolated Herbig Ae/Be stars; MWC 297

{\bf High mass star forming regions} ~~ NGC 2024 and NGC 2068 clusters in
Orion; Monoceros R2 cloud; Sharpless 106; W3 B ultracompact H{\sc II}
regions; NGC 3603 young star cluster; Rosette Nebula and molecular
cloud; M 16 Eagle Nebula; M 17 Omega Nebula; Sgr C cloud; Galactic
Center mosaic; many studies of extragalactic H{\sc II} regions and
star clusters in nearby galaxies

{\bf OB and Wolf-Rayet stars} ~~ $\tau$ Sco; $\delta$ Ori; $\lambda$ Ori;
$\iota$ Ori; HD 206267; $\zeta$ Pup; $\eta$ Carina; $\gamma^2$ Vel; WR
140; WR 147

\section{Synthesis of findings}

The table below provides a broad-brush summary of the findings outlined 
above.  In the nearer star forming regions where molecular clouds are 
small and few high-mass stars, the X-ray luminosity is modest ($L_x 
\simeq 10^{31}-10^{33}$ erg/s and is a produced by a combination of 
magnetic activity in lower-mass stars and stellar winds in higher-mass 
stars.  The more distant giant molecular clouds with rich high-mass 
young stellar clusters have X-ray luminosities of order $10^{33}-
10^{36}$ erg/s where the principal contributors are Wolf-Rayet stars 
(binaries?) and diffuse emission from individual and merged supernova 
remnants.  In starburst environments that involve many molecular clouds 
and star formation over $10^7$ years or longer, the X-ray luminosity is 
in the range $10^{36}-10^{41}$ erg/s where the emission is mainly 
produced by X-ray binaries and an interstellar medium superheated by 
many supernova remnants.

\begin{table}
\caption{Overview of Early Chandra Results on Star Forming Regions}
\begin{tabular}{cccccccc}
Region    & Size & Total  &  \multicolumn{2}{c}{Indiv.\ Stars} &&
\multicolumn{2}{c}{Diffuse}\\ \cline{4-5} \cline{7-8}
          & (pc) & log$L_x$& log$L_x$ & Type && log$L_x$& Type \\
&&&&&&& \\
Ophiuchi  &   1 & 32 & 28-31 & B$-$M stars    &&none &         \\
          &     &    &       & + protostars   &&     &         \\
&&&&&&& \\
NGC 1333  &   1 & 31 & 28-30 & O$-$M stars    && none &         \\
&&&&&&& \\
Orion     &   1 & 33 & 28-31 & 0.05-50 M$_\odot$ && none & \\
&&&&&&& \\
Sgr A*    &  10 & 36 & 33    & XRBs           && 34 & SN remnant \\  
\\
region    &     &    & 32    & WR stars       && 36 & Superhot ISM \\
&&&&&&& \\
Sgr B2    &  10 & 34 & 33    & cluster        && 34 & Fluorescence \\
&&&&&&& \\
30 Dor    & 100 & 36 & 33-34 & $> 100 {\rm M}_\odot$&& 36 & SNRs \\
&&&&&&& \\
M 82      &1000 & 40 & 38-39 & XRBs           && 40& Superhot ISM \\
\tableline
\tableline
\end{tabular}
\end{table}

The X-ray properties of star forming regions are thus quite complex.  
When the star formation rate is high enough to produce many OB stars, 
star birth and star death are spatially intertwined.  But in these 
massive star forming regions, the X-ray emission from the death phases 
(supernova remnants, X-ray binaries) typically outshine the emission 
from the birth phases (protostar, T Tauri, Herbig Ae/Be, OB stars) by a 
large factor.  Pre-main sequence and OB stellar X-ray emission is 
undoubtedly present  at low levels in many fields of supernova remnants 
and starburst regions. 

\section{Astrophysical issues}

A wide range of astrophysical issues arise from X-ray studies of young
stars and star forming regions.  Some involve the mechanisms of X-ray
emission: Exactly how do OB winds produce X-rays?  Do low-mass pre-main
sequence stellar magnetic fields arise from a dynamo, as in the Sun, or
in other ways? Are protostellar X-rays produced in solar-type magnetic
fields rooted in the stellar surface, or in field lines reaching from
the star to the disk?

But the implications of young stellar X-ray emission that may have the 
most profound implications involve the ionization effects of the X-rays 
on ambient molecular material (Glassgold et al.\ 2000).  X-ray energy 
is typically deposited at column densities around $10^{21} < \log N_H < 
10^{23}$ cm$^{-2}$ from the source.  Thus every embedded protostar, T 
Tauri star and OB star will produce an X-ray dissociation region (XDRs) 
of low-ionization fraction (Hollenbach \& Tielens 1997).  These are 
have much lower ionization fraction but extend to larger distances than 
Stromgren spheres and photodissociation regions around embedded OB 
stars.  X-ray ionization will dominate over cosmic ray ionization that 
is thought to uniformly permeate molecular clouds: these XDRs may 
extend $10^{-2}$ pc around typical T Tauri stars and $10^0$ pc around 
X-ray-luminous OB stars (Lorenzani \& Palla 2001; these are shown as 
grey regions in Figure 1).  Elevated ionization of largely neutral 
molecular gas will increase the coupling between the gas and magnetic 
fields by slowing ambipolar diffuse.  This may have a critical 
inhibiting effect on future star formation;  see Bertoldi \& McKee 
(1996) for the theory of photoionization self-regulation of star 
formation.  

On smaller scales, X-ray ionization is definitely likely to be the 
principal ionization source within protostellar systems. X-ray 
ionization can penetrate deeply into circumstellar disks, stimulating 
MHD instabilities and accretion, and is probably important for the 
coupling between disks and bipolar outflows (Glassgold et al.\ 2000).  

\acknowledgements
I thank the many Chandra scientists who shared ideas and results
reported in this talk: Gordon Garmire, Richard Griffiths, Nicholas
Grosso, Kensuke Imanishi, Katsuji Koyama, Yoshitomo Maeda, Thierry
Montmerle, Hiroshi Murakami, Francesco Palla, Steven Pravdo, Norbert 
Schulz, Leisa Townsley, Yohko Tsuboi.  This work was supported by NASA 
contract NAS8-38252.

\end{document}